\documentclass{jfm}

\usepackage{graphicx}
\usepackage{newtxtext}
\usepackage{newtxmath}
\usepackage{natbib}
\usepackage{hyperref}
\usepackage{xcolor}
\hypersetup{
    colorlinks = true,
    urlcolor   = blue,
    citecolor  = black,
}

\newcommand{\RomanNumeralCaps}[1]
\linenumbers

\newcommand\MC[1]{{\color{black}#1}}

\title{Prandtl number effects on extreme mixing events in forced stratified turbulence}

\shorttitle{Prandtl number effects on stratified turbulence}

\author{Nicolaos Petropoulos\aff{1}
  \corresp{\email{np546@cam.ac.uk}}, Miles M. P. Couchman\aff{2}, Ali Mashayek\aff{3}, Stephen M. de Bruyn Kops\aff{4} \and Colm-cille P. Caulfield\aff{5,1}}

\affiliation{\aff{1}Department of Applied Mathematics and Theoretical Physics, University of Cambridge, Cambridge CB3 0WA, UK
\aff{2}Department of Mathematics and Statistics, York University, Toronto, ON M3J 1P3, Canada
\aff{3}Department of Earth Sciences, University of Cambridge, Cambridge CB2 3EQ, UK
\aff{4}Department of Mechanical and Industrial Engineering, University of Massachusetts Amherst, Amherst,
MA 01003, USA
\aff{5}Institute for Energy and Environmental Flows, University of Cambridge, Cambridge CB3 0EZ, UK
}

\begin{document}
\maketitle

\begin{abstract}
Relatively strongly stratified turbulent flows tend to self-organise into a  `layered anisotropic stratified turbulence' (LAST) regime, characterised by relatively deep and well-mixed density `layers' separated by relatively thin `interfaces' of enhanced density gradient. 
Understanding the associated mixing dynamics is a central problem in geophysical fluid dynamics. It is challenging to study `LAST' mixing, as it is associated with Reynolds numbers $Re := UL/\nu \gg 1$ and Froude numbers $Fr :=(2\pi U)/(L N) \ll 1$, ($U$ and $L$ being characteristic velocity and length scales, $\nu$ being the kinematic viscosity and $N$ the buoyancy frequency). Since a sufficiently large dynamic range (largely) unaffected by stratification and viscosity is required, it is also necessary for the buoyancy Reynolds number $Re_{b} := \epsilon/(\nu N^{2}) \gg 1$ where $\epsilon$ is the (appropriately volume-averaged) turbulent kinetic energy dissipation rate. This requirement is exacerbated for oceanically relevant flows, as the Prandtl number $Pr := \nu/\kappa = \mathcal{O}(10)$ in thermally-stratified water (where $\kappa$ is the thermal diffusivity), thus leading (potentially) to even finer density field structures. We report here on four forced fully resolved direct numerical simulations of stratified turbulence at various Froude ($Fr=0.5, 2$) and Prandtl numbers ($Pr=1, 7$) forced so that $Re_{b}=50$, with resolutions up to $30240 \times 30240 \times 3780$.
We find that, as $Pr$ increases, emergent `interfaces' become finer and their contribution to bulk mixing characteristics decreases at the expense of the small-scale density structures populating the well-mixed `layers'. However, extreme mixing events (as quantified by  significantly elevated local destruction rates of buoyancy variance $\chi_0$) are always preferentially found in the (statically stable) interfaces, irrespective of the value of $Pr$.
\end{abstract}



\section{Introduction}
\label{section:introduction_Prandtl}

In density-stratified flows, small-scale turbulence is known to enhance the rate at which density gradients are irreversibly smoothed by diffusive processes (i.e. mixed) and quantifying this rate in geophysical, environmental and industrial settings is of crucial importance. For instance, turbulent mixing is known to have a leading order impact on global oceanic circulations \citep{wunsch2004vertical}. 
Stratified turbulent flows are characterised by a variety of length scales associated with the structure of the turbulent velocity and density fields. Velocity fluctuations are smoothed by molecular viscosity and dissipated into heat below the Kolmogorov scale $L_{K} := (\nu^{3}/\epsilon)^{1/4}$ (where $\nu$ is the molecular viscosity and $\epsilon$ is the appropriately volume-averaged dissipation rate of turbulent kinetic energy). Similarly, density fluctuations are dissipated by molecular diffusion below the Batchelor scale $L_{\rho} := (\nu\kappa^{2} / \epsilon)^{1/4}$, $\kappa$ being the molecular diffusivity. 
The (molecular) Prandtl number $Pr := \nu / \kappa$, that effectively quantifies the relative strength of viscous dissipation to diffusion, is the ratio of the square of these two length scales. Unlike the atmosphere, in which $Pr \simeq 0.7$ ($L_K \approx L_{\rho}$), in the ocean $Pr \sim \mathcal{O}(10)$ when thermally stratified and the equivalent Schmidt number $Sc:=\nu/D \sim \mathcal{O}(1000)$ ($D$ being the salt diffusivity) when dominantly salt-stratified.

For turbulent flows with $Pr \gg 1$, the inevitable scale separation with $L_\rho \ll L_K$  is such that fully resolved numerical simulations are highly challenging. However, there is increasing evidence indicating that the value of $Pr$ has (perhaps unsurprisingly) a leading order impact on irreversible scalar-mixing properties of stratified turbulent flows. For instance, the effects of $Pr$ variations on the properties of secondary instabilities arising from the breakdown of Kelvin-Helmholtz billows were reported by \citet{mashayek2011turbulence} and \citet{salehipour2015turbulent}. Similarly, \citet{zhou2017self} studied the influence of $Pr$ variations on fully developed turbulence in stratified plane Couette flows using direct numerical simulations (DNS) and reported significant effects on density and momentum fluxes. Also using DNS, \citet{legaspi2020prandtl} analysed the effects of $Pr$ variations on homogeneous forced stratified turbulence and showed, at least for the range of parameters they considered, that the $Pr=1$ simulations are able to describe $Pr > 1$ dynamics at large scales and also that kinetic energy spectra (which importantly do not directly contain information from the density field), are largely unaffected by variations in $Pr$. However, they did observe $Pr$ effects at small scales below $L_K$ and in the density flux spectra. Therefore, here we focus on the influence of $Pr$ variations on small-scale structures of the density field and their associated effects on the mixing properties of forced stratified turbulent flows. 

Another important length scale associated with stratified turbulent flows is the buoyancy length scale $L_{B} := 2\pi U / N_{0}$ where $U$ is a characteristic velocity of the flow and $N_{0}$ is a characteristic (background) value of the buoyancy frequency. Scaled using a characteristic horizontal length scale $L_{h}$, this parameter then defines a horizontal Froude number $Fr_{h} := L_{B} / L_{h} = 2\pi U / N_{0}L_{h}$ that quantifies the relative strength of the stratification of the flow. For small horizontal Froude numbers $Fr_{h}$ (i.e. relatively strongly stratified flows), the density field is known to self-organise into relatively well-mixed `layers' (whose size scales as $L_{B}$; the buoyancy scale can therefore be thought of as the largest energetically possible overturn) separated by relatively thin `interfaces' of enhanced density gradient \citep{ billant2001self,waite2011stratified}. Such density `staircase' structure has important implications for irreversible scalar mixing in density-stratified turbulent flows. 
\citet{couchman2023mixing} showed that (in the $Pr=1$ case) while static instabilities are most prevalent within the well-mixed layers (that are hence characterised by relatively high values of $\epsilon$), much of the scalar mixing, as described by the (appropriately volume-averaged) buoyancy variance destruction rate $\chi$, is located in the relatively strongly stratified interfaces, a phenomenon that is not apparent when considering $\epsilon$ only. Hence,  we focus here on the contribution of the different structures described above to the overall mixing properties of the flow, as described by $\chi$. 

Fundamentally, we aim to understand how the density field's small-scale organisation shapes the bulk mixing properties of forced stratified turbulent flows at different Prandtl numbers. To this end, we analyse fully resolved DNS data of forced stratified turbulent flows at $Pr=1$ and $Pr=7$, each with $Fr=0.5$ and $Fr=2$, for values of $Re$ such that the emergent $Re_{b}=50$. The rest of this paper is organised as follows. We  describe the DNS data in section~\ref{section:summary_DNS} and then present the methodology used to extract distinct classes of density field structures in section~\ref{section:methodology}. In section~\ref{section:cluster_properties}, we apply this methodology to the DNS data to segment the density field into weakly and strongly stratified interfaces, relatively small-scale `lamella-like' structures and larger scale density inversions 
and analyse the associated mixing properties of the extracted density structures. Brief conclusions are drawn in section~\ref{section:discussion_Prandtl}. 

\section{Summary of the DNS datasets}
\label{section:summary_DNS}

\begin{table}
  \begin{center}
\def~{\hphantom{0}}
  \begin{tabular}{cccccccccccc}
    Name  & $Pr$   &   $Fr$ & $Re$ & $Re_\lambda$ & Domain size & Grid size & $L_{K}$ & $L_{T}$ & $L_{B}$ & $\epsilon $ & $ \chi $ \\
    P1F200 & 1 & 2 & 1717 & 184 & $2\pi \times 2\pi \times \pi$ & 4096 $\times$ 4096 $\times$ 2048 & 0.0037 & 0.099 & 1.99 & 1.05 & 0.54 \\
    P1F050 & 1 & 0.5 & 24096 & 826 & $2\pi \times 2\pi \times \pi/4$ & 11264 $\times$ 11264 $\times$ 1408 & 0.00052 & 0.030 & 0.52 & 0.95 & 0.38 \\
    P7F200 & 7 & 2 & 1717 & 178 & $2\pi \times 2\pi \times \pi$ & 6144 $\times$ 6144 $\times$ 3072 & 0.0037 & 0.096 & 1.95 & 1.09 & 0.44 \\
    P7F050 & 7 & 0.5 & 24096 & 655 & $2\pi \times 2\pi \times \pi/4$ & 30240 $\times$ 30240 $\times$ 3780 & 0.00048 & 0.024 & 0.51 & 1.36 & 0.43 \\
  \end{tabular}
  \captionsetup{width=1\linewidth}
  \caption{Summary of the DNS data. $L_{K}$, $L_{T}$ and $L_{B}$ denote the Kolmogorov, Taylor and buoyancy scales. 
  }
  \label{tab:parameter_space}
  \end{center}
\end{table}


We consider statistically steady, forced, fully-resolved DNS of stratified turbulence from the simulation campaign originally reported by~\citet{almalkie2012kinetic}. The non-hydrostatic dimensionless Navier-Stokes equations under the Boussinesq approximation,
\begin{equation}
    \begin{cases}
        \partial_{t}\boldsymbol{u} + \boldsymbol{u}\bcdot\bnabla\boldsymbol{u} = - \bnabla p + \frac{1}{Re}\nabla^{2}\boldsymbol{u} -\left(\frac{2\pi}{Fr}\right)^{2}\rho\hat{\boldsymbol{z}} + \mathcal{F}, \quad \bnabla \bcdot \boldsymbol{u} = 0, \\
        \partial_{t}\rho + \boldsymbol{u} \bcdot \bnabla \rho = \frac{1}{Pr Re}\nabla^{2}\rho, 
    \end{cases}
  \end{equation}
  are numerically integrated using a pseudospectral code in a triply periodic domain, where
  $\hat{\boldsymbol{z}}$ is the (upward) vertical unit vector. The dimensionless parameters are:
  the above-defined Prandtl number $Pr$;
  the Froude number $Fr := 2\pi U/(N_{0}L)$ (where $U$ and $L$ are characteristic velocity and length scales associated with the forcing and $N_{0}$ is the background buoyancy frequency); and the Reynolds number $Re := UL/\nu$. The forcing term $\mathcal{F}$ corresponds to the deterministic `Rf' scheme described in~\cite{rao2011mathematical} and is designed to match a target low wavenumber kinetic energy spectrum at steady-state. This ensures a buoyancy Reynolds number $Re_{b} :=  \epsilon / (\nu N_{0}^{2}) \simeq 50$ for these simulations, where $\epsilon $ is the volume-average of the point-wise local  dissipation rate of turbulent kinetic energy $\epsilon_0$, defined in terms of the symmetric part of the strain-rate tensor $s_{ij}$ as:
  \begin{equation}
    \epsilon_0 := \frac{2}{Re} s_{ij} s_{ij} ; \ s_{ij} := \frac{1}{2} \left ( \frac{\partial u_i}{\partial x_j}+ \frac{\partial u_j}{\partial x_i} \right ).
    \label{eq:eps0def}
    \end{equation}
The density field is a superposition of a background linear density field $\overline{\rho}(z)$, characterised by a reference density $\rho_{0}$ and reference density gradient $-N_{0}^{2}\rho_{0}/g$,  and a perturbation $\rho^{\prime}(\boldsymbol{x}, t)$ so that, in dimensional form,
\begin{equation}
    \rho(\boldsymbol{x}, t) = \overline{\rho}(z) + \rho^{\prime}(\boldsymbol{x}, t) := \rho_{0}(1-N_{0}^{2}z/g) + \rho^{\prime}(\boldsymbol{x}, t). 
  \end{equation}
  We can use $\rho^{\prime}$ to compute the (dimensional) local destruction rate of buoyancy variance: 
\begin{equation}
    \chi_0 := \frac{g^{2}\kappa}{\rho_{0}^{2}N_{0}^{2}}\bnabla \rho^{\prime} \bcdot \bnabla \rho^{\prime}. 
  \end{equation}
This quantity, appropriately scaled in this way by $-\rho_{0}N_{0}^{2} / g$ (i.e. the density gradient against which the turbulence acts),  corresponds to the local destruction rate of available potential energy (density) and can therefore be used as a proxy of local irreversible mixing~\citep{howland2021quantifying} which can still be calculated pointwise-locally in the flow domain.  Henceforth, we consider dimensionless quantities (the prefactor being $1/(Pr Re)$ in our system) and denote by $\chi $ the volume-average of local $\chi_0$ across the entire computational domain. 

Here, we consider two values of the Prandtl number ($Pr =\left\{ 1,7\right\}$) as well as two values of the Froude number ($Fr = \left\{0.5, 2\right\}$). The simulation parameters are summarised in table~\ref{tab:parameter_space}. We choose the grid spacing $\Delta \lesssim L_{\rho}$, and a vertical domain dimension of approximately $1.5L_{B}$. We consider a single snapshot in time (at statistical steady-state) of the various simulations. 

\begin{figure}
    \captionsetup{width=1\linewidth}
    \centering
    \includegraphics[width=\linewidth]{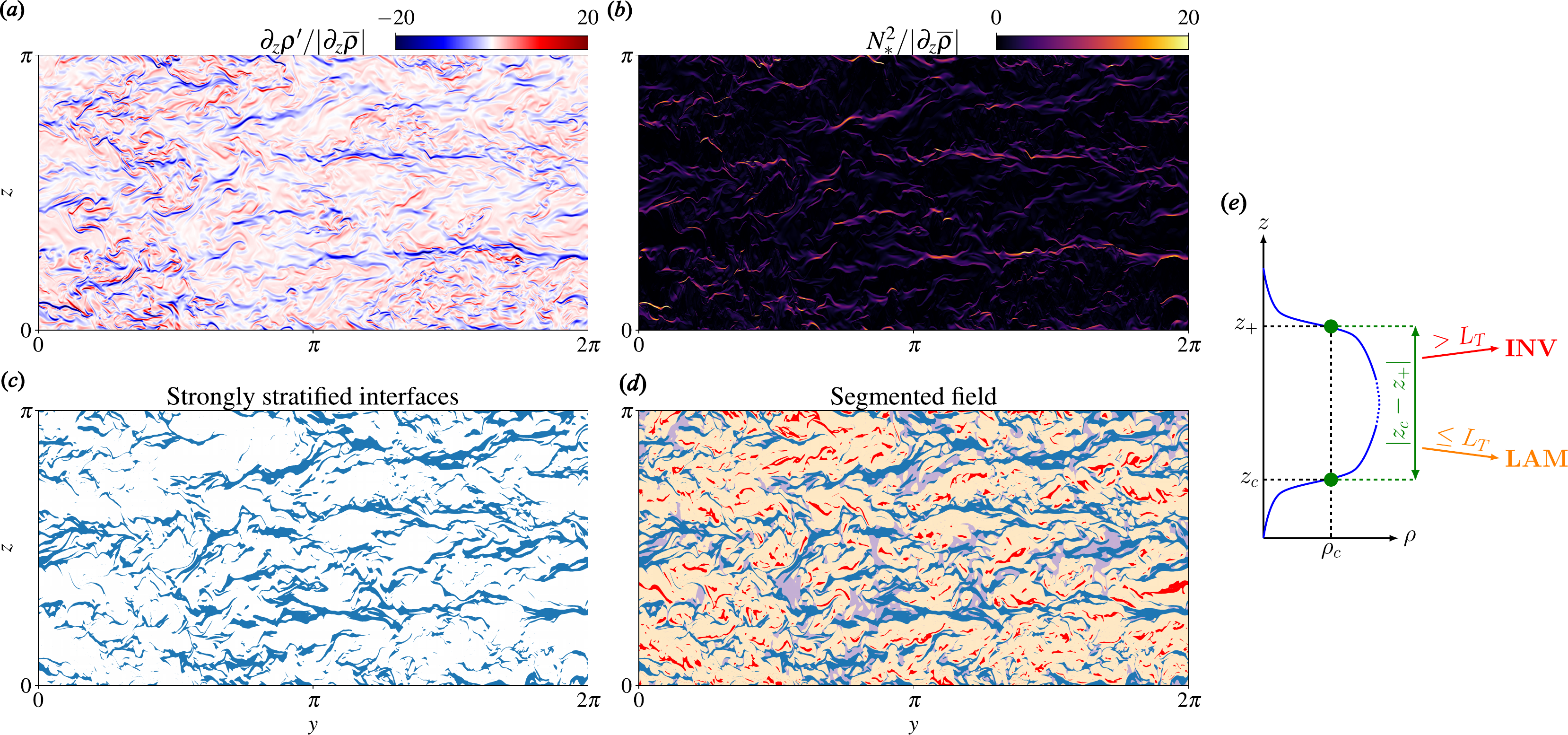}
    \caption{Flow segmentation methodology. For the P1F200 simulation, the density field, represented here by its vertical gradient as shown in panel $(a)$, is vertically sorted into a statically stable field, whose vertical gradient is denoted $N^{2}_{\ast}$ (up to a multiplicative constant), as shown in panel $(b)$. The sorting algorithm highlights the stable interfaces of the density field, i.e. the points of the density field unaltered by the sorting procedure. The strongly stratified interfaces (\textbf{SINT}, in blue with $N^{2}_{\ast}>1$) are then extracted in panel $(c)$. The entire segmented field is shown in panel $(d)$,
      including the weakly stratified interfaces with $N^{2}_{\ast} \leq 1$  (\textbf{WINT}, in purple), and 
the relatively well-mixed regions between interfaces, further sub-divided into small-scale `lamella' structures (\textbf{LAM}, in orange) and larger-scale density inversions (\textbf{INV}, in red) using the procedure described in section~\ref{section:methodology} and shown schematically in panel $(e)$, based on the Taylor microscale $L_T$.}
    \label{fig:methodology}
\end{figure}

For the relatively weakly stratified $Fr=2$ simulations, a patch of elevated turbulence, described by relatively high values of the local dissipation rate  $\epsilon_0$, is generated by a relatively large-scale vertically aligned vortex such as the one reported in~\citep{couchman2023mixing} for the P1F200 dataset. 
Outside the vortex, the density field self-organises into relatively well-mixed layers separated by interfaces characterised by enhanced density gradients (see figures~\ref{fig:methodology} and~\ref{fig:segmented_fields}). In the relatively strongly stratified $Fr=0.5$ case, the vortex does not appear. 
\citet{couchman2023mixing} showed (at least for the P1F200 simulation) that the interfaces account for relatively large values of $\chi_0$ and low values of $\epsilon_0$ and hence `extreme' values of the (local) flux coefficient $\Gamma_0 := \chi_0 / \epsilon_0$, whereas the well-mixed layers were potentially more turbulent (relatively large values of $\epsilon_0$) but characterised by relatively low values of $\chi_0$ (owing to relatively low values of the density gradient) and hence relatively low values of $\Gamma_0$. Here we aim to understand whether such a picture holds for flows with different values of $Fr$ and more importantly of $Pr$, and understand how different structures in the density field (defined more precisely in the next section) influence the mixing within the flow, quantified by values of (local) $\chi_0$. 


\section{Density field segmentation methodology}
\label{section:methodology}

\begin{figure}
    \captionsetup{width=1\linewidth}
    \centering
    \includegraphics[width=\linewidth]{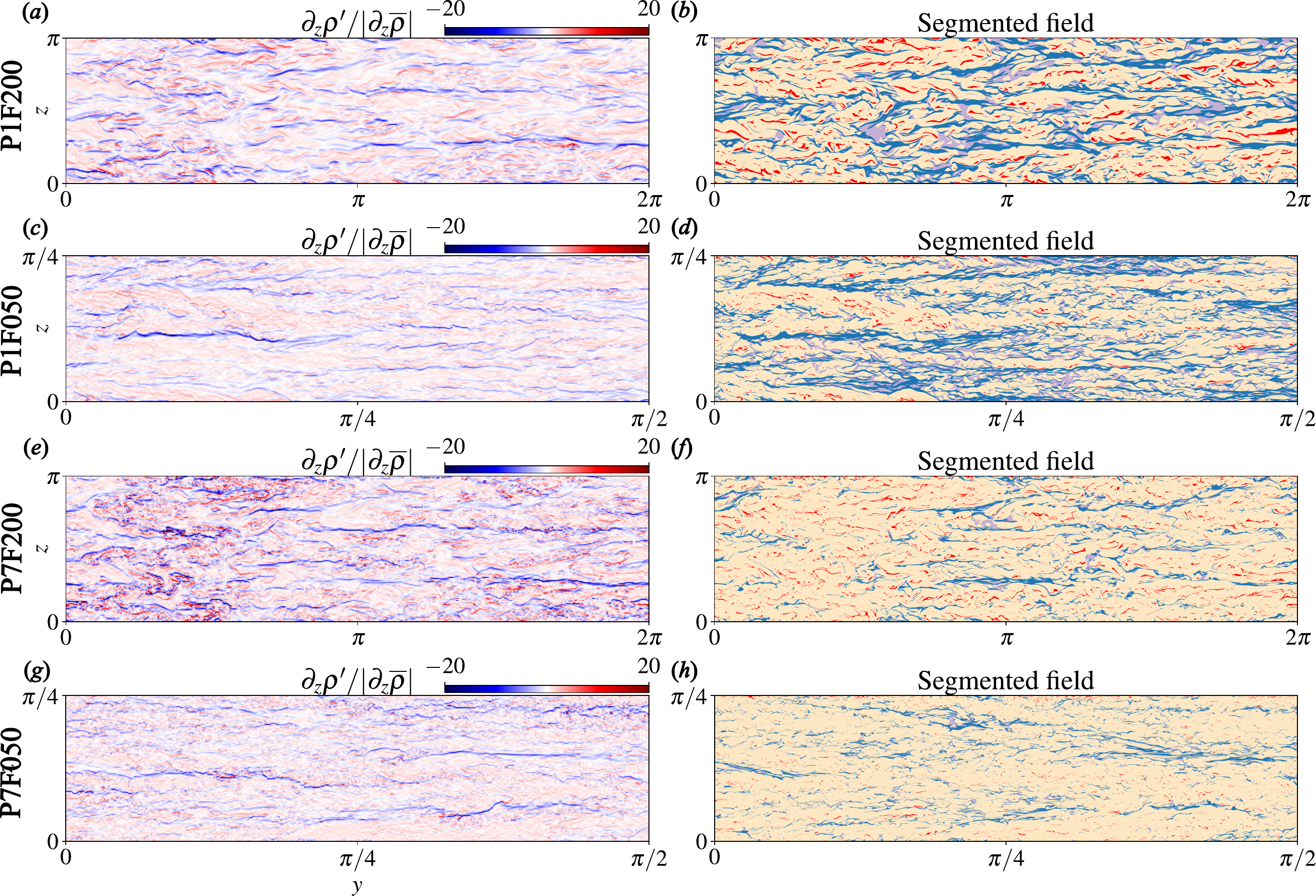}
    \caption{Normalized vertical density gradient $\partial_{z}\rho^{\prime}/\vert \partial_{z}\overline{\rho} \vert$ (panels $(a,c,e,g)$) and associated segmented fields (panels $(b,d,f,h)$), for simulation P1F200 ($a,b$), P1F050 ($c,d$), P7F200 ($e,f$) and P7F050 ($g,h$). The strongly stratified interfaces (\textbf{SINT}) are depicted in blue, the small-scale structures (\textbf{LAM}) in orange, the larger-scale density inversions (\textbf{INV}) in red and the weakly stably stratified regions (\textbf{WINT}) in purple. }
    \label{fig:segmented_fields} 
\end{figure}

Following \cite{couchman2023mixing}, we are interested in the contribution to mixing of the emergent stably stratified interfaces appearing in the density field and how this contribution varies as a function of $Pr$ and $Fr$. As $Pr$ increases, we expect finer structures to arise in the intervening well-mixed regions and we also are interested in understanding how these structures shape the mixing properties of the flow. Therefore, prior to any analysis, the flow is segmented into four different categories based on the local (vertical) density gradient $\partial_{z}\rho$ as well as the neighbouring structure of the flow. We apply the following methodology, as summarized in figure~\ref{fig:methodology}: 
\begin{itemize}
    \item {\it Stably stratified interfaces}: We sort vertical density profiles by values of the density $\rho$ in order to create statically stable density profiles $\rho^{\ast}(\boldsymbol{x})$. We identify points of the dataset whose values of the density remain unaltered by the sorting procedure, i.e. points that lie in statically stable regions of the dataset. These points have relatively high values of the background buoyancy gradient $N_{\ast}^{2} := - g/(\rho_{0}N_{0}^{2})\partial_{z} \rho^{\ast}$ (see figure~\ref{fig:methodology}(\emph{b-c})) and correspond to stably stratified interfaces lying between relatively well-mixed regions, such as those reported by~\citet{couchman2023mixing} for the P1F200 dataset.  We thus refer to such points as belonging to the \textbf{INT} (`interface') cluster, which is further subdivided into relatively strongly stratified (\textbf{SINT}) and relatively weakly stratified (\textbf{WINT}) interfaces, for $N_{\ast}^{2} > 1$ and $N_{\ast}^{2} \leq 1$ respectively. 
    \item {\it Relatively well-mixed layers}: We subdivide points that are altered by the sorting procedure (i.e. are not within an interface \textbf{INT})  into two categories depending on the local density gradient $\partial_{z} \rho$ and neighbouring structure of the (unsorted) density field $\rho$,
in particular relative to $L_T$, the Taylor microscale of the flow. $L_T$ describes the scale at which viscosity starts to affect the development of turbulent eddies significantly, defined (dimensionally) as $L_T:= \sqrt { 10 \nu {\mathcal K} /\epsilon}$, where ${\mathcal K}$ is the volume averaged turbulent kinetic energy.
      As illustrated in figure \ref{fig:methodology}(e), by considering a point in the dataset at vertical coordinate $z_{c}$ and density $\rho_{c}$ satisfying $\partial_{z} \rho(z_{c})> 0$, we define $z_{+}$ as the closest point moving upwards for which $\rho(z_{+}) = \rho_{c}$. As equality is difficult to ensure numerically, we define
    \begin{equation}
        z_{+} := \text{min}\{z > z_{c} \; \vert \; \rho(z) \leq \rho(z_{c})\}. 
    \end{equation}
    Similarly, if $\partial_{z} \rho(z_{c}) < 0$, we define $z_{-} := \text{min}\{z < z_{c} \; \vert \; \rho(z) \geq \rho(z_{c})\}$.
    If $\lvert z_{c} - z_{+} \rvert \leq L_{T}$ (or $\lvert z_{c} - z_{-} \rvert \leq L_{T}$ depending on the local value of the density gradient), 
    we have identified a relatively small scale structure in the density field, such as a blob or stretched `lamella' as discussed in detail by \cite{villermaux2019mixing}, which is strongly affected by viscosity. Therefore we classify these structures as belonging to the \textbf{LAM} (`lamella') cluster. Conversely, if $\lvert z_{c} - z_{+} \rvert > L_{T}$ (or $\lvert z_{c} - z_{-} \rvert > L_{T}$), we have identified a point belonging to a relatively large scale density inversion largely unaffected by viscosity, 
    and so we classify it as being in the \textbf{INV} (`inversion') cluster.
    This segmentation methodology, summarised in figure~\ref{fig:methodology}, is applied to the four datasets considered in this work, as shown in  figure~\ref{fig:segmented_fields}.

\end{itemize}

\section{Cluster properties}
\label{section:cluster_properties}

\subsection{Contributions to $\chi$}

\begin{figure}
    \captionsetup{width=1\linewidth}
    \centering
    \includegraphics[width=\linewidth]{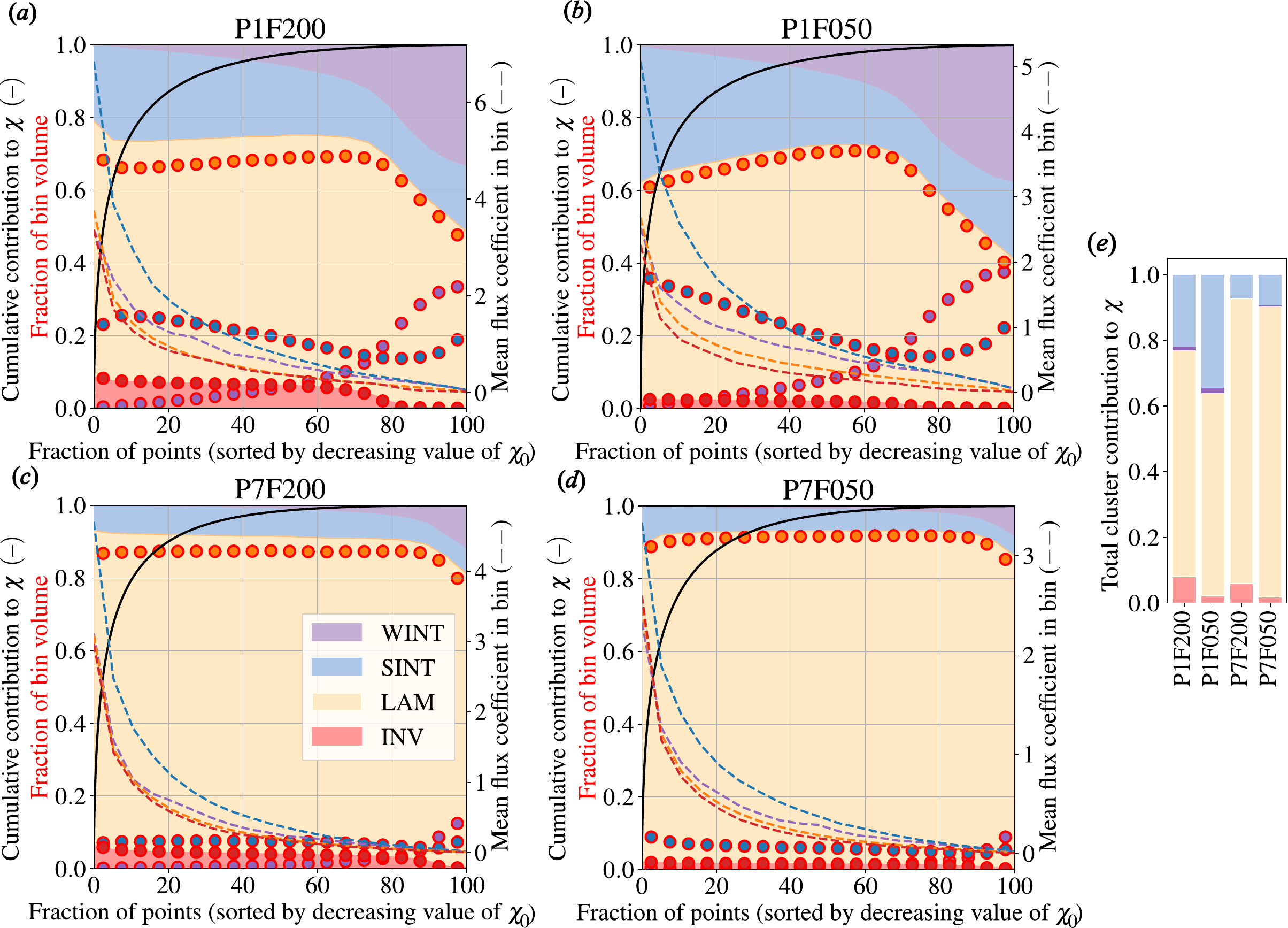}
    \caption{$(a$-$d)$ Normalized cumulative contribution to $\chi $ (black line; see equation~\eqref{eq:cumulative_contribution}) for each simulation. Data points are assigned to 20 equal volume bins, sorted by decreasing $\chi_0$ and clustered using the method presented in section~\ref{section:methodology}. For each bin, we compute the {\it relative} contribution of each cluster to $\langle \chi \rangle_{\text{bin}}$, as shown by the heights of the colored regions. For each bin, we compute the fraction of points within each cluster to the bin's total number of points (red circled dots), along with the (arithmetic) mean value of $\Gamma_0 := \chi_0 / \epsilon_0$ for each cluster  (dashed lines). (\emph{e}) Total contributions to $ \chi $ from each cluster for the four simulations.  }
    \label{fig:results}
\end{figure}

We consider the relative importance of each cluster defined in section~\ref{section:methodology} in terms of their contribution to the local and bulk mixing properties of the flow, as described by local $\chi_0$ and volume-averaged $\chi$. We particularly focus on the role of `extreme' mixing events and therefore sort the data by decreasing values of $\chi_0$, defining a sorted vector $\boldsymbol{\chi}_0^{*} = (\chi_{0}^{0*}, \cdots, \chi_{0}^{M*})$ where $M$ is the number of points in the dataset and $\chi_{0}^{0*}$ and $\chi_{0}^{M*}$ are the greatest and lowest values of $\chi_{0}$ found in the dataset, respectively. This vector can be used to construct the normalized cumulative contribution to $\chi$ as
\begin{equation}
    \forall n \in \{1, \cdots, M\}, \; \chi^{c}(n) := \frac{1}{\chi}\sum_{i=1}^{n}\chi_{0}^{i*}. 
    \label{eq:cumulative_contribution}
\end{equation}
The cumulative contribution $\chi^{c}$ is plotted in figure~\ref{fig:results} for each dataset (solid black line), highlighting the fact that the dominant contribution to $\chi$ is generated by a relatively small set of localised `extreme' events, regardless of $Fr$ and $Pr$.   More specifically, for the four sets of parameters considered in this work, approximately $80\%$ of the contribution to $\chi $ is contained within the first $10\%$ of the points sorted by decreasing values of $\chi_0$ (and hence $10\%$ of the box volume). Similar statistics are observed in oceanographic data~\citep{couchman2021data}. 

The {\it total} contribution of each cluster to bulk $\chi$ is shown in figure~\ref{fig:results}(\emph{e}). We also assign the data points sorted by values of $\chi_0$ into 
 $n=20$ equal volume bins, in order to calculate a {\it relative} contribution of points in each cluster to the bin-volume-averaged  $\langle \chi \rangle_{\text{bin}} := \sum_{\text{bin}}\chi_0/(V/n)$, where $V$ is the total number of points in the domain, as illustrated by colored shading in figures~\ref{fig:results}(\emph{a-d}). 
 Focusing on the $Pr = 1$ case, the contribution (figure~\ref{fig:results}(\emph{e})) from the strongly stratified interfaces (\textbf{SINT} cluster) to bulk $\chi$ is roughly $25-35\%$, whereas the contribution from small-scale structures (\textbf{LAM}) reaches $\sim 60\%$. As  $Fr$  decreases and stratification strengthens, the total contribution from large-scale inversions (\textbf{INV}) shrinks from $\sim 10\%$ for $Fr=2$ to less than $3\%$ for $Fr=0.5$, possibly due to the
suppression of large-scale overturnings by
 relatively strong stratification. We observe similar trends for the relative contributions in each bin. 
In increasing the Prandtl number from $Pr=1$ to $7$, strong interfaces become (perhaps unsurprisingly) finer (figure~\ref{fig:segmented_fields}(\emph{f,h})) and their total contribution shrinks to about $10\%$, whereas the total contribution from small-scale structures in `lamella' (\textbf{LAM}) increases to about $80-90\%$. The total contribution from  large-scale inversions does not exceed $10\%$. Again, we observe similar trends in the relative contributions as $\chi_0$ decreases. 

\begin{figure}
    \captionsetup{width=1\linewidth}
    \centering
    \includegraphics[width=\linewidth]{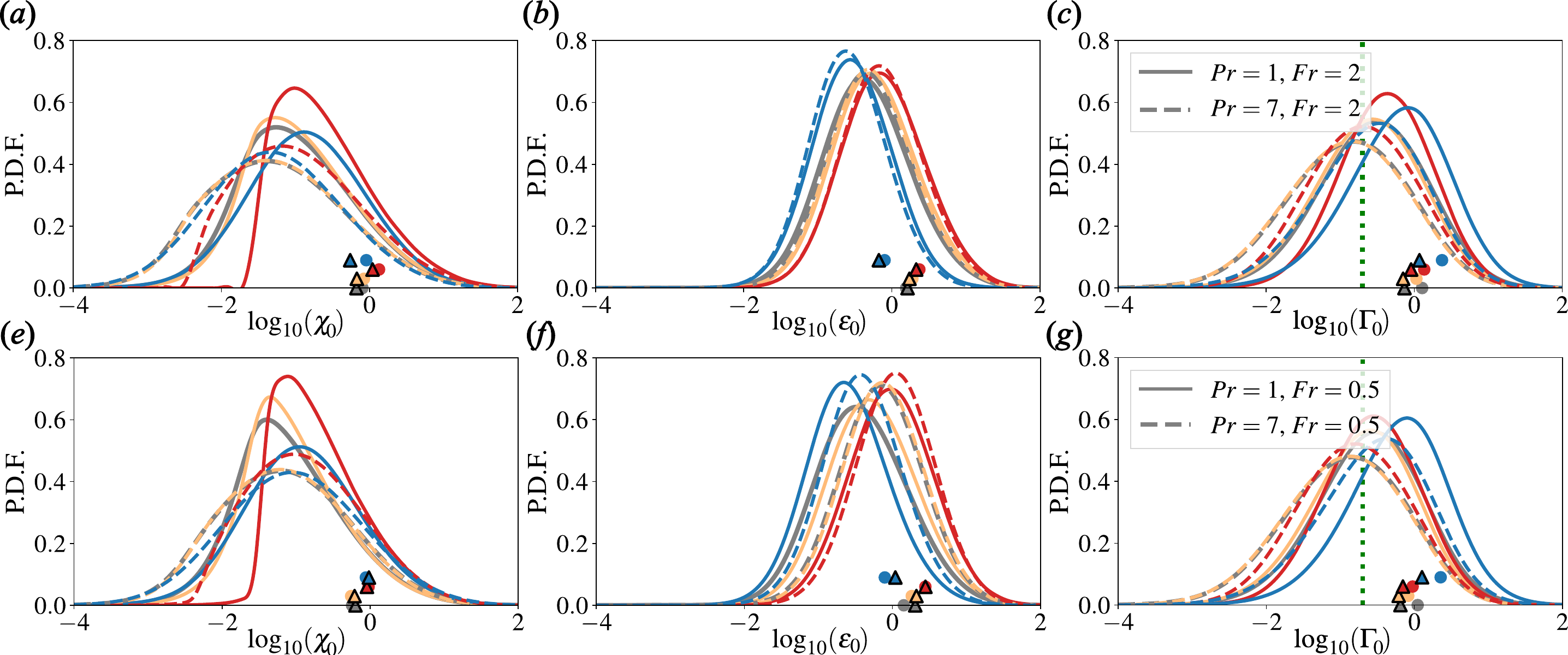}
    \caption{Probability density functions (PDF) for $\log_{10}(\chi_0)$ (panel (\emph{a}), (\emph{e})), $\log_{10}(\epsilon_0)$ (panel (\emph{b}), (\emph{f})) and $\log_{10}(\Gamma_0)$ (panel (\emph{c}), (\emph{g})) for the different clusters defined in section~\ref{section:methodology} and for the four simulations. The statistical mean of each field (without the logarithm) is represented by a colour-coded circle ($Pr=1$) or triangle ($Pr=7$). The green dotted vertical line corresponds to the canonical value $\Gamma=0.2$. }
    \label{fig:conditional_PDF}
\end{figure}

\subsection{Statistics of $\Gamma$}

For each bin in figure~\ref{fig:results}, we compute a mean value of the (local) flux coefficient $\Gamma_0 := \chi_0 / \epsilon_0$ associated with each cluster (see dashed lines), although such mean values of ratios of local quantities should always be treated with caution.  The leftmost bins (corresponding to the most `extreme' values of $\chi_0$) have the largest values of $\Gamma_0$ (of order unity, significantly above the canonical value $\Gamma = 0.2$) suggesting strongly that  the extreme mixing events (large $\chi_0$) are not necessarily correlated with extreme turbulent events (large $\epsilon_0$). Among the extreme events in $\chi_0$, those in the strongly stratified interfaces (\textbf{SINT}) cluster correspond to the largest mean values of $\Gamma_0$ (as suggested by~\citet{couchman2023mixing} for the $Pr=1$, $Fr=2$ case), independently of variations in $Pr$ and $Fr$. This point is further emphasized in figure~\ref{fig:conditional_PDF} where we plot the probability density functions (PDFs) of $\log_{10}(\chi_0)$, $\log_{10}(\epsilon_0)$ and $\log_{10}(\Gamma_0)$  for the different clusters and simulations. We also present the statistical mean of $\chi_0$, $\epsilon_0$ and $\Gamma_0$. For each simulation, the statistical mean of the flux coefficient $\Gamma_0$ is maximized for the strongly stratified interfaces (almost twice as large as the statistical mean for the entire dataset), because of relatively low values of $\epsilon_0$ and large values of $\chi_0$. As the Prandtl number increases, the statistical mean of $\Gamma_0$ decreases (a result also observed by~\citet{salehipour2015turbulent}) for both the weakly and strongly stratified cases, for the following (different) reasons.

In the weakly stratified simulations with $Fr=2$ (figures~\ref{fig:conditional_PDF}(\emph{a-c})), density effectively acts as a passive scalar (as can be seen from the weakly stratified scaling of~\citet{riley1981direct}, for instance) and hence the statistics of $\epsilon_0$ do not depend on $Pr$, as can be seen in figure~\ref{fig:conditional_PDF}(\emph{b}). However, as $Pr$ increases, the left tails of the $\chi_0$ distributions become more significant, as might be expected for the mixing of an effectively passive scalar, as  $\chi_0$ is multiplied by $(PrRe)^{-1}$ where $Re$ is fixed. We note that a widening of the tails of $\partial_{z}\rho^{\prime}$ (not shown here, see for instance~\citet{Riley_2023}) effectively rebuilds the right tail of $\chi_0$, explaining why the PDF of $\chi_0$ is not just shifted toward lower values. A more in-depth discussion of this phenomenon is provided by \citet{bragg2023understanding}.
As a result \MC{of the statistics of $\epsilon_0$ remaining roughly independent of $Pr$ but those of $\chi_0$ decreasing with increasing $Pr$}, the statistics of $\Gamma_0$ decrease as $Pr$ increases for the weakly stratified ($Fr=2$) case. 

Conversely, in the strongly stratified simulations with $Fr=0.5$, buoyancy acts `actively' on the momentum field, as demonstrated for instance by the strongly stratified scaling analysis of~\citet{billant2001self}. Moreover, as $Pr$ increases, the volume contribution of the small-scale structures (\textbf{LAM} cluster) increases at the expense of the strongly stratified interfaces (\textbf{SINT}) (see figure~\ref{fig:results}). \textbf{LAM} structures, which locally are only weakly affected by stratification because they populate the relatively well-mixed regions of the flow, are viscously affected (as their vertical extent is, by definition, smaller than $L_T$) but still significantly disordered. As a result, their local static instability inevitably encourages {\it increased} \MC{viscous} dissipation, and so their enhanced prevalence actually shifts the statistics of $\epsilon_0$ towards higher values for the $Pr=7$ flows as compared to $Pr=1$ \MC{(yellow curves, figure~\ref{fig:conditional_PDF}(\emph{f}))}. 
In this sense, enhanced stratification at higher $Pr$ actually makes the flow `more turbulent', reminiscent of the prediction by \citet{pearson1983} of relatively long-lived `approximately horizontal striations' in high-$Pr$ decaying stratified turbulence. Also, though this is a second order effect, for the strongly stratified simulations the left tail of the $\chi_0$ distribution once again becomes somewhat more significant as $Pr$ increases. 
\MC{We note that while the forcing scheme endeavors to maintain a constant $Re_b$, bulk $\epsilon$ is slightly mismatched between the two $Pr$ simulations at $Fr=0.5$ (see Table 1). Therefore, while increased $Pr$ dramatically changes the relative prevalence of \textbf{LAM} versus \textbf{SINT} structures, part of the observed increase in local $\epsilon_0$ statistics at $Pr=7, Fr=0.5$ is due to the slight mismatch in targeted bulk $\epsilon$ rather than being a purely $Pr$ effect.} 
Importantly, it is apparent in figure~\ref{fig:conditional_PDF}(\emph{e}) that the total amount of irreversible mixing for the flow with $Pr=7$ actually slightly \emph{increases} \MC{compared to $Pr=1$} in the more strongly stratified $Fr=0.5$ simulations, essentially because the $Pr=7$ flow is more vigorous, \MC{despite the fact that $\Gamma_0$ appears to decrease with increasing $Pr$. Therefore,} consideration of $\Gamma_0$, or even  $\Gamma:=\chi/\epsilon$ (constructed from volume-averaged dissipation rates), in isolation should be treated with caution.

\section{Discussion}
\label{section:discussion_Prandtl}
We have analysed the influence of the Prandtl number $Pr$ and Froude number $Fr$ on density structures in forced turbulent stratified flows and their contribution to irreversible scalar mixing properties. Using fully resolved DNS data and a flow segmentation algorithm based on the local value of the (vertical) density gradient and the local (vertical) structure of the density field, we have extracted distinct regions of the turbulent density field \---  interfaces (both relatively strong and relatively weak)  separating well-mixed density layers made up of both small-scale lamellar structures and larger scale density inversions \--- and analysed their contribution to the bulk value of the destruction rate of buoyancy variance $\chi$ and to the statistics of the (locally evaluated) flux coefficient $\Gamma_0$. 

As $Pr$ increases, the strongly stratified density interfaces become finer and their contribution to average values of $\chi$ decreases at the expense of the small-scale lamellar structures in the relatively well-mixed density layers. However, similarly to the flow with $Pr=1$ \citep{couchman2023mixing}, these structures are `quiescent',  yet mixing hotspots. The points in these structures associated with the largest values of (local) $\chi_0$ are associated with extreme values of $\Gamma_0$ and hence with relatively low values of $\epsilon_0$. More generally, the flux coefficient associated with strongly stratified structures is  (in an averaged sense) essentially twice as large as its value for the three other segmented regions, regardless of the values of $Pr$ and $Fr$. All in all, strongly stratified interfaces are therefore characterised by relatively large values of $\chi_0$ and relatively weak values of $\epsilon_0$ compared to the other density structures considered in this work and are therefore likely to be overlooked if considering $\epsilon_0$ (or indeed the volume average $\epsilon$) alone as a proxy for significant irreversible mixing. 

As is becoming increasingly well-appreciated, a universal value of $\Gamma$ is not able to capture the complex and inhomogeneous structures of the density and density gradient turbulent fields. Moreover, our description of the density fields as well as consideration of the considered mixing regime (i.e. `passive' mixing at higher $Fr$ as compared to  `active' mixing at lower $Fr$) offers a potential explanation for the empirical observation that the `mixing efficiency' $\Gamma/(1+\Gamma)$  decreases with $Pr$. On the one hand, mixing in the relatively weakly stratified case can be described as `passive': the velocity field and hence $\epsilon_0$ are largely unaffected by $Pr$ but the small-scale structure of the density field evolves as $Pr$ increases from $1$ to $7$ in a way that results in a decrease in \MC{$\chi_0$ and thus the} flux coefficient $\Gamma_0$. On the other hand, mixing in the relatively strongly stratified case can be thought of as `active', with the density field and its small-scale lamellar structure having a leading order impact on the rate at which the turbulent kinetic energy is dissipated \MC{($\epsilon_0$)}, ultimately leading to a decrease in $\Gamma_0$ as $Pr$ increases. This shows another way in which a focus on the flux coefficient can be misleading. Since the decrease in $\Gamma_0$ is actually principally related to an increase in $\epsilon_0$, the total amount of mixing (quantified by $\chi$) actually \emph{increases} in the higher $Pr$, strongly stratified flow considered here. Another key result of our analysis is that the Prandtl number $Pr$ has a leading order impact (at least compared to the Froude number $Fr$) on the density field structures and their mixing properties, further emphasizing the need to take this parameter into account when `measuring mixing', a process that is inherently a diffusive one~\citep{villermaux2019mixing, Annurev_Colm_Caulfield} and that affects the density field. 

We here considered steady-state forced stratified turbulent flows, and so it would now be interesting to apply our segmentation and associated mixing contribution analysis to time-dependent decaying stratified turbulent flows. For example,
what is the fate of the different emerging structures? Can a separate analysis of the temporal evolution of each structure help us better understand the mixing history of a stratified turbulent flow as a whole and unveil the fundamental rules of stratified mixing? These are questions left for future work. 



\backsection[Funding]{This project received funding from the European Union's Horizon 2020 research and innovation program under the Marie Sklodowska-Curie Grant Agreement No. 956457 and used resources of the Oak Ridge Leadership Computing Facility at the Oak Ridge National Laboratory, supported by the Office of Science of the U.S. Department of Energy under Contract No. DE-AC05-00OR22725. S.deB.K. was supported under U.S. Office of Naval Research Grant number N00014-19-1-2152. For the purpose of open access, the authors have applied a Creative Commons Attribution (CC BY) licence to any Author Accepted Manuscript version arising from this submission. }





\bibliographystyle{jfm}
\bibliography{jfm}


\end{document}